\documentclass[12pt,preprint]{aastex}
\newcommand{\twcott}{$^{12}$CO $J$=3$-$2}
\newcommand{\twcoto}{$^{12}$CO $J$=2$-$1}
\newcommand{\twcooz}{$^{12}$CO $J$=1$-$0}
\newcommand{\twco}{$^{12}$CO}
\newcommand{\thcott}{$^{13}$CO $J$=3$-$2}
\newcommand{\thcoto}{$^{13}$CO $J$=2$-$1}
\newcommand{\thcooz}{$^{13}$CO $J$=1$-$0}
\newcommand{\thco}{$^{13}$CO}

\newcommand{\jtt}{$J$=3$-$2}
\newcommand{\jto}{$J$=2$-$1}
\newcommand{\joz}{$J$=1$-$0}
\newcommand{\etc}{etc}
\newcommand{\etal}{et al.}
\newcommand{\eg}{e.g.}

\newcommand{\kms}{km s$^{-1}$}

\newcommand{\tmb}{$T_{\rm MB}$}
\newcommand{\msun}{M$_\odot$}
\newcommand{\h}{$^{\rm h}$}
\newcommand{\m}{$^{\rm m}$}

\shorttitle{CO in Double-Barred Galaxies}
\shortauthors{Petitpas \& Wilson}

\begin{document}

\title{Molecular Gas in Candidate Double-Barred Galaxies II. Cooler, Less Dense Gas
Associated with Stronger Central Concentrations}

\author{Glen R. Petitpas}
\affil{University of Maryland}
\affil{College Park MD, USA 20742}
\email{petitpas@astro.umd.edu}
\and
\author{Christine D. Wilson}
\affil{McMaster University}
\affil{1280 Main Street West, Hamilton ON,
Canada L8S 4M1}
\email{wilson@physics.mcmaster.ca}

\begin{abstract}

We have performed a multi-transition CO study of the centers of seven
double-barred galaxies that exhibit a variety of molecular gas
morphologies to determine if the molecular gas properties are
correlated with the nuclear morphology and star forming activity.
Near infrared galaxy surveys have revealed the existence of nuclear
stellar bars in a large number of barred or lenticular galaxies. High
resolution CO maps of these galaxies exhibit a wide range of
morphologies. Recent simulations of double-barred galaxies suggest
that variations in the gas properties may allow it to respond
differently to similar gravitational potentials. We find that the
\twcott/\jto\ line ratio is lower in galaxies with centrally
concentrated gas distributions and higher in galaxies with CO emission
dispersed around the galactic center in rings and peaks. The
\thco/\twcoto\ line ratios are similar for all galaxies, which
indicates that the \jtt/\jto\ line ratio is tracing variations in gas
temperature and density, rather than variations in optical
depth. There is evidence that the galaxies which contain more
centralized CO distributions are comprised of molecular gas that is
cooler and less dense. Observations suggest that the star formation
rates are higher in the galaxies containing the warmer, denser, less
centrally concentrated gas. It is possible that either the bar
dynamics are responsible for the variety of gas distributions and
densities (and hence the star formation rates) or that the star
formation alone is responsible for modifying the gas properties.

\end{abstract}

\keywords{Galaxies: starburst -- galaxies: active -- galaxies: ISM --
galaxies: nuclei -- galaxies: individual (NGC 470, NGC 1097, NGC 2273,
NGC 3081, NGC 4736, NGC 5728, NGC 6951)}

\section{Introduction \label{intro}}

Recent near infrared (NIR) surveys reveal isophote twists in the
central regions of barred galaxies, which are thought to be the
signature of a bar within a bar \citep[\eg,][]{mul97}. Recent models
of double-barred galaxies have had some success in reproducing
relatively long-lived central features that can explain the isophote
twists.  \citet{sha93} model galaxies as gravitational stars and
dissipative gas clouds and find that the NIR isophote twists are
caused by viscous and gravitational torques that drag the central
regions of the main bar out of alignment with the rest of the
bar. \citet{fri93} suggest that the NIR isophote twists are the result
of a kinematically distinct ``nuclear bar'' that can rotate with up to six
times the pattern speed of the large-scale bar. One thing these models
have in common is the need for dissipation. Both groups have to
include large amounts of molecular gas in order for the models to be
able to reproduce and sustain the observed features. The models of
barred galaxies by \citet{com85}, \citet{sha93}, and \citet{fri93}
produce very different nuclear morphologies by varying the amount of
gaseous component included in the simulations. More problematically, the
models of \citet{com94} and \citet{mac02} exhibit different central
morphologies by simply changing the {\it properties} of the
interstellar medium.

In an attempt to test these models, we have undertaken a study of the
molecular gas properties of a sample of galaxies believed to contain
double bars based on the NIR isophote twists observed in their nuclei
(see Tables \ref{galaxytable} and \ref{nucleustable}).
In Paper I \citep{pet02} we find that, despite similarities in the NIR
images, the CO maps show very different morphologies. For example, the
CO maps of NGC 2273, NGC 2782, and NGC 4736 exhibit what may be
interpreted as nuclear bars that align with the nuclear NIR isophote
twists \citep{pet02,jog99,won00}, while NGC 6951 shows twin peaks of
molecular gas collecting at an inner Lindblad resonance
\citep{koh99}. NGC 5728 shows a very disordered CO structure where the
clumps of CO emission do not seem to align with any structures seen at
other wavelengths \citep{pet02}.
 
Simulations of galaxies (both single- and double-barred) have shown that
the details of the nuclear structure and inflow rates depend on the
assumed gas properties of the models. For example, \citet{com94}
demonstrated that the nuclear bar of a double-barred galaxy can either
be phase locked \citep[as predicted by][]{sha93} or kinematically
separated from the primary bar \citep[as predicted by][]{fri93}
depending on the gas viscosity in the models. More recently,
\citet{mac02} modeled the flow of gas in a realistic double-barred
potential and found that the inflow rates depend on the sound speed of
the interstellar medium. Galaxies with high sound speeds had greater
inflow rates and the gas was able to reach all the way into the
nucleus.  In the galaxies with lower sound speeds, the infalling gas
is trapped into a circumnuclear ring near the Inner Lindblad Resonance
(ILR) and even the torques exerted by the nuclear stellar bar are not
enough to transport the gas inwards beyond the ring. This model
suggests that nuclear bars may not be an effective means of fueling
Active Galactic Nuclei (AGN) as originally proposed by \citet{shl89}.

While the current generation of millimeter and submillimeter
telescopes are not capable of resolving molecular gas on the parsec
scales needed to resolve the AGN fueling mechanisms directly, we can
study the gas within the regions of the ILR. In this paper we present
a detailed set of molecular gas data for a sample of double-barred
galaxies. We have chosen 7 galaxies that are known to contain nuclear
stellar bars based on their NIR images \citep[see for
example,][]{sha93,fri96,jun97,mul97,woz95}, of which 5 have high
resolution \twcooz\ data published. The properties of these galaxies
are summarized in Table \ref{galaxytable}. This sample will allow us
to determine if variations in molecular gas properties are linked in
some way to the wide variety of CO morphologies in a sample of
galaxies with similar NIR morphologies, as hinted at by the models of
\citet{com94} and \citet{mac02}.

In \S\ref{obs} we discuss the observations and data reduction
techniques. In \S\ref{phys} we determine the molecular gas physical
condition using the Local Thermodynamic Equilibrium (LTE) and Large
Velocity Gradient (LVG) approximations. In \S\ref{morph} we discuss
observed correlations between our CO line ratios with the high
resolution CO morphology and simulations of double-barred galaxies. In
\S\ref{disc1} we speculate on possible origins of the observed
correlations. This work is summarized in \S\ref{conc}.

\section{Observations and Data Reduction \label{obs}}

The nuclei of seven double-barred galaxies were observed using the
James Clerk Maxwell Telescope (JCMT)\footnote{The JCMT is operated by
the Joint Astronomy Centre in Hilo, Hawaii on behalf of the parent
organizations Particle Physics and Astronomy Research Council in the
United Kingdom, the National Research Council of Canada and The
Netherlands Organization for Scientific Research.} over the period of
1998 - 2001. We have obtained \twcoto, \thcoto, and \twcott\
detections for all galaxies listed in Table \ref{galaxytable} except
\thcoto\ in NGC 3081. The half-power beam width of the JCMT is $21''$
at 230 GHz (\twcoto) and $14''$ at 345 GHz (\twcott). \twcott\ data
were obtained on a five-point cross offset by $7''$ in order to allow
convolution to match the 21$''$ beam of the JCMT at 230/220 GHz. All
observations were obtained using the Digital Autocorrelation
Spectrometer in beam switching mode with a 180\arcsec\ throw. The
calibration was monitored by frequently observing spectral line
calibrators. The spectral line calibrators showed very little scatter
from the published values with individual measurements differing by
typically $< 15\%$ from standard spectra. Thus, we adopt the nominal
main beam efficiencies ($\eta_{\rm MB}$) from the JCMT Users Guide of
0.69 at 230/220 GHz and 0.63 at 345 GHz.

Initially, similar data sets were averaged together using the software
package SPECX. Spectra were output to FITS files and then the data
were further reduced using the Bell Labs data reduction package
COMB. The data were binned to 10 km s$^{-1}$ resolution (7.8 and 11.5
MHz at 230 and 345 GHz respectively) and zeroth or first order
baselines were removed. The emitting regions were quite wide ($>$ 200
km s$^{-1}$) but the spectrometer bandwidth was at least 800 km
s$^{-1}$, which allowed for accurate baseline determination. The
exception is NGC 5728 which had lines that were $>$ 450 \kms\ wide and
for which the \twcoto\ spectrum ran off the low velocity end of the
spectrometer. For this reason, only the high velocity continuum was
used in baseline removal for this galaxy. Finally, the five \twcott\
spectra for each galaxy were convolved to simulate a $21''$
beam. The spectra for each galaxy are shown in Figure \ref{spectra},
and the spectral line intensities are summarized in Table
\ref{spectratable}.

The data were then binned to 20 \kms\ resolution. This allowed the
calculation of line ratios for each channel within the spectral line.
The ratios for each channel were averaged 
using equal weights to obtain the final line
ratios. Only those channels with a signal-to-noise ratio of at least
two in the line ratio were used in the calculation. The
uncertainty in the line ratio is taken as the standard deviation of
the mean. As a final step, the spectra and line ratios were scaled to
the main beam temperature scale using the appropriate values for
$\eta_{\rm MB}$ (see above). A summary of the line ratios is given in
Table \ref{ratiostable} and channel by channel histograms of the
ratios are shown in Figure \ref{ratios}. We note that the \twcoto\
spectral line for NGC 5728 is truncated somewhat (see Figure
\ref{spectra}). We feel that the truncation of the \jto\ spectra will
have less effect on our CO line ratios, since they were calculated
channel by channel, rather than integrated intensity over the entire
spectral line.

The channel-by-channel \twcott/\jto\ line ratios for NGC 1097 and NGC
6951 shown in Figure \ref{ratios} show evidence for gradients. This
may be caused by pointing offsets in the observations of one
transition relative to the other, or may reflect a true gradient in
excitation temperature across the nuclei of these galaxies. The
individual line profiles for the central positions are similar in the
\twcott\ and \twcoto\ data for these galaxies, suggesting that large
pointing offsets are not present.  Evidence for similar gradients has
been seen in the nearby starburst galaxy M82 \citep{pet00} and may be
caused by variation in star formation strengths across the nuclei.  In
the case of NGC 1097, the slight gradient could easily be produced by
slight pointing offsets, but the gradient in NGC 6951 is strong enough
that it may reflect true variations in excitation temperature across
the nucleus of this galaxy. In either case, we are averaging the gas
properties across the entire nucleus, so the gradients (real or
artificial) will not significantly affect our results.

For \S\ref{lvg}, the Large Velocity Gradient models contain four free
parameters, so we need more line ratios in order to place stronger
constraints on the molecular gas physical conditions. We have searched
the literature and have obtained \twcooz\ line strengths for all
galaxies except NGC 3081. All \twcooz\ spectra were taken with the
Nobeyama Radio Observatory (NRO) with 16\arcsec\ resolution except for
NGC 2273 which was taken at the IRAM 30 m telescope with a
22\arcsec\ beam (see Table \ref{spectratable} for references). Since
our JCMT \twcoto\ data are taken at 21\arcsec\ resolution, we have
created beam-matched \twcott/\joz\ ratios with the JCMT and NRO data and
divided them by our beam-matched \twcott/\jto\ ratio. In the case of
NGC 4736, we used the \twcoto/\joz\ line ratio taken at IRAM with a
22\arcsec\ beam by \citet{ger91} (since they did not publish the
individual line strengths). For the line ratios we took from the
literature or created using JCMT and published data, we adopt a 30\%
uncertainty to account for cross calibration uncertainties.

\section{Molecular Gas Physical Conditions \label{phys}}

All of these galaxies were initially chosen because the NIR images
suggested the presence of a nuclear stellar bar. The similarities in
the NIR images suggest that the gravitational potentials are similar
in these galaxies (see Table \ref{nucleustable}). Models of
double-barred galaxies indicate that the gas may behave differently
within similar potentials if the gas properties are varied. In this
section, we will model the CO line ratios to determine the true
physical conditions of the molecular gas, such as excitation
temperature, kinetic temperature, density, and column density.

Determining the physical conditions of molecular clouds in other
galaxies can be very difficult. For single dish observations, the beam
size is usually larger than the angular diameter of typical molecular
clouds, and so the line strengths are beam-diluted. In addition,
molecular clouds are known to be very clumpy \citep[\eg,][]{stu90} and
the high resolution \twcooz\ maps of the galaxies in our sample show a
variety of distributions within the inner 21\arcsec\
\citep{ken92,jog98,pet02,sak99}, so the average emission likely
originates in a mixture of high and low density material with
different beam filling factors. For extragalactic sources, the filling
factor of the CO emission within the primary beam may be small and is
difficult to determine accurately. Thus, for low resolution, single
dish studies such as this one, it is preferable to study the ratios of
line strength, rather than the individual fluxes.  As long as the two
CO transitions are observed with the same size primary beam, the
division of the two eliminates the unknown filling factor of the
emission within the beam of the telescope (assuming the two
frequencies originate from the same region, which is likely true for
these distant sources). 

The high resolution CO maps of these galaxies suggest that most of the
emission originates from the regions at or interior to the ILR (as
traced by star forming rings or the ends of nuclear bars). The
exception is NGC 4736 for which at 4 Mpc the 21\arcsec\
primary beam of the JCMT does not cover the entire ILR region
\citep[][see Figure \ref{radial}]{won00}.  In all galaxies though, the
JCMT beam entirely covers the strongest emission regions, which
suggests that the physical conditions determined here represent a good
approximation of the {\it average} gas properties within the ILR
regions.

In \S\ref{lte} we will analyze the line ratios under the Local
Thermodynamic Equilibrium (LTE) approximation. In this model (in
addition to the unknown filling factor, $f_\nu$), assumptions have to be
made about the optical depth of the emission lines. In \S\ref{lvg} we
will analyze the line ratios under the Large Velocity Gradient (LVG)
assumption, which is more complicated but does not require any
assumption regarding optical depths.

Calibration differences between different telescopes can introduce
(often large) uncertainties in spectral line ratios \citep[see
discussion in][for example]{til91}. To prevent the introduction of such
uncertainties, we will limit our analysis in \S\ref{lte} to the
interpretation of only those CO line ratios taken with the JCMT, namely
\thco/\twcoto\ and \twcott/\jto. In the LVG approximation, we will need
more than two CO line ratios to place strong constraints on the
molecular gas conditions, so in \S\ref{lvg} we will use all of the line
ratios listed in Table \ref{ratiostable}.

\subsection{LTE Analysis \label{lte}}

In the LTE approximation, we assume that the gas is collisionally
dominated and hence the relative populations between molecular energy
levels are a function of a single temperature. 
There are generally three basic limits with which we are concerned:
\begin{enumerate}
\item{
Both lines are optically thick ($\tau \gg 1$), in which case the
integrated intensity line ratio (\eg\ \twcott/\jto) can be written as
\begin{equation}
R_{32/21}^{\rm thick} = {[J_{\nu_{32}}(T_{32}) - J_{\nu_{32}}(T_{\rm bg})]
\over{[J_{\nu_{21}}(T_{21}) - J_{\nu_{21}}(T_{\rm bg})] }} \label{thickratio}
\end{equation}
where $J_\nu(T)$ is proportional to the Planck function
\citep[see][]{kut81} and $T_{\rm bg}= 2.7$ K. In LTE ($T_{32} = T_{21}
= T_{\rm ex}$) $R_{32/21} \approx 1$ for $T > 30$ K (see Figure
\ref{lteplots}).  }
\item{
Both lines are optically thin ($\tau \ll 1$) and the integrated
intensity line ratio can be written as 
\begin{equation}
R_{32/21}^{\rm thin} = {[J_{\nu_{32}}(T_{32}) - J_{\nu_{32}}(T_{\rm bg})]\tau_{32}
\over{[J_{\nu_{21}}(T_{21}) - J_{\nu_{21}}(T_{\rm bg})]\tau_{21} }}. \label{thinratio}
\end{equation}
which in LTE depends on the optical depth at each frequency,
$\tau_\nu$. The ratio of $\tau$'s can be expressed as a function of
$J$, $\nu$, CO-specific constants, and the Boltzmann equation
\citep[see for example][]{sch87}.  }
\item{
One line is optically thick and the other is optically thin, as is
likely the case when we compare integrated intensity line ratios for
different isotopomers of CO in the same $J$ transition
(\eg\ \thco/\twcoto). If we assume $T_{ex}^{\rm ^{13}CO} = T_{ex}^{\rm
^{12}CO}$ and that \twcoto\ is optically thick and \thcoto\ is
optically thin, we have
\begin{equation} 
R_{12/13} \approx {X\over{\tau ({\rm ^{12}CO})}}
\end{equation}
where $X$ is the \twco/\thco\ abundance ratio \citep[which is measured
to be 43-62 in the Milky Way;][]{haw87,lan93}.
}
\end{enumerate}

Plots of the integrated intensity line ratio as a function of
excitation temperature for the optically thick and thin cases are
given in Figure \ref{lteplots}. The excitation temperatures derived
from the \twcott/\jto\ line ratio under the assumption of LTE are
summarized in Table \ref{ltesols}.

The \twco/\thcoto\ line ratio is uniform at the 2$\sigma$ level over
all galaxies and has an average value of 0.11. Assuming optically
thick \twco\ and optically thin \thco\ emission, the \thco/\twcoto\
line ratio suggests that the \twco\ optical depth in all six galaxies
is consistent with $\tau \approx 5$. This indicates that the molecular
gas is on the border between optically thin and optically thick in all
of our galaxies.

In the LTE approximation, we are unable to determine densities
accurately, but we can say that, given the similarities in optical depth
between the galaxies, the higher \twcott/\jto\ line ratios indicate
warmer gas. We will discuss the implications of this in later
sections.

\subsection{LVG Analysis \label{lvg}}

As a double check of the LTE analysis, we have performed a LVG
analysis using a code written by Lee Mundy and implemented as part of
the MIRIAD data reduction package. A sample of the output from the
models is shown in Figure \ref{lvgplots}. The integrated intensity
line ratios are shown as bands which indicate the $\pm$30\% upper and
lower limits of the line ratios. We adopt 30\% to account for the
uncertainties involved in spectral line calibration and baseline
removal as well as, in the case of the \twcoto/\joz\ line ratio, the
uncertainties involved in combining spectral lines from different
telescopes. Solutions are indicated by the locations where the regions
of the different ratio boundaries overlap. In a few cases, only one of
the line ratio contours is visible on the plot, so arrows indicate the
allowed region. A grayscale representation of the \twcott/\jto\ line
ratio is included to aid interpretation of this figure. A single
component model provided an adequate fit to the observed line ratios
in most cases.

There are four main variables in the LVG models: CO column density per
unit velocity ($N$(CO)$/dv$), molecular hydrogen density ($n$(H$_2$)),
kinetic temperature ($T_{\rm kin}$), and the \twco/\thco\ abundance
ratio ([\twco]/[\thco]). With only three line ratios, we cannot place
very strong constraints on the physical conditions of the molecular
gas. We have performed a study of parameter space ranging from 15 $<$
log $N$(CO)/dv (cm$^{-2}$/\kms) $<$ 21, 1 $<$ log $n({\rm H}_2)$
(cm$^{-3}$) $<$ 6, 30 $<$ [\twco]/[\thco] $<$ 70, and 30 K $< T_{\rm
kin} <$ 100 K. In general, solutions were found for all temperatures
and abundance ratios. Changes in the [\twco]/\thco] ratio made
negligible changes to the solution compared to changes in the kinetic
temperature. For this reason, we adopt [\twco]/[\thco] = 50,
consistent with the value obtained for the Milky Way
\citep{haw87,lan93}. The derived values from the LVG analysis are
summarized in Table \ref{lvgsols}. For the purpose of clarity, we show
results only for $T_{\rm kin} = 30$ K (consistent with the dust
temperatures determined by the IRAS fluxes, see Table \ref{ltesols}).

In many cases we can only determine lower limits to the molecular
hydrogen density, and thus we have upper limits on the optical depth.
It is interesting that all galaxies have rather similar column
densities of CO per unit velocity (within the uncertainty). They are
generally all $\sim 10^{17}$ cm$^{-2}$(\kms)$^{-1}$. If we wish to
convert this to a true column density of CO, we must multiply by the
velocity width of a typical molecular cloud in each galaxy.
Unfortunately, the large beam of the JCMT covers many individual clouds
in these distant galaxies in a single pointing, so the line width of
individual clouds is difficult to determine. In the nearest galaxy in
our sample (NGC 4736) our single dish CO spectra shows emission
features (\eg\ $V_{LSR} = 225$ \kms; Figure \ref{spectra}) with line
widths of $\sim$ 50 \kms. High resolution \twcooz\ maps of these
galaxies indicate line widths of $\sim$30-50 \kms\ \citep{koh99}. Even
these are still prone to the effects of galaxy rotation. For
consistency, we will assume that individual clouds in all galaxies have
similar CO line widths of 10 \kms, similar to the Milky Way.

In the case of NGC 6951, our JCMT line ratios do not overlap in Figure
\ref{lvgplots}. Thus we cannot place strong constraints on the gas
properties in the center of this galaxy. The values listed in Table
\ref{lvgsols} are determined using only the \twcott/\jto\ line ratio,
and are not as firm as those determined for the other galaxies.

Adopting a line width of 10 \kms\ gives CO column densities of 10$^{18}$
cm$^{-3}$ for most of the galaxies in our sample. In order for star
formation to occur, it is believed that the molecular gas must
exceed some critical column density. This value is believed to be on
the order of 10$^{22}$ cm$^{-3}$ \citep{sol87,elm89}. Assuming a number
density ratio of $n$(CO)/$n$(H$_2$) = 10$^{-4}$ \citep{gen92} we find a
column density of molecular hydrogen of 10$^{22}$ cm$^{-3}$ which may
explain why we see such a high level of nuclear activity in these
galaxies.

\subsection{Overall Gas Properties from LTE and LVG}

The $J=1$, $J=2$, and $J=3$ levels of CO lie at 5.5 K, 17 K,
and 33 K above the ground state, respectively. In addition, in the
optically thin limit where all CO emission escapes the cloud, the
critical densities of H$_2$ needed to collisionally excite \twco\ at
the same rate that it spontaneously decays are $\sim$ 10$^3$, 10$^4$,
and 10$^5$ cm$^{-3}$ for the \joz, \jto, and \jtt\ transitions,
respectively. So, fundamentally, higher \twcott/\jto\ line ratios
suggest that the molecular gas is warmer and/or denser. 

In the simpler LTE approximation, we cannot accurately determine the
density of the gas, but higher \twcott/\jto\ line ratios suggest
warmer gas in these galaxies. We do not have enough line ratios to
place strong constraints on the gas properties using the LVG models
but, in general, higher \twcott/\jto\ line ratios also indicate higher
temperature (for a fixed density). Conversely, for a fixed temperature
(which is feasible considering the similarities of the dust
temperatures in these galaxies) the higher \twcott/\jto\ line ratios
indicate denser gas. So in either the LVG or LTE model, the higher
\twcott/\jto\ line ratio suggests warmer and/or denser gas.

The similarities in the \twco/\thcoto\ line ratios predict optical
depths of $\sim 5$ in all galaxies in both the LTE and LVG
approximations \citep[assuming Galactic values for the \twco/\thco\
abundance ratio of $\sim 50$;][]{haw87,lan93}. This similarity in
optical depths makes the conclusions regarding temperature and density
drawn from the \twcott/\jto\ line ratio more convincing because we do
not have to worry as much about variations in optical depth modifying
the line ratio.

\section{A Correlation of CO Line Ratios with Central CO Morphology \label{morph}}

The existence of nuclear stellar bars in these galaxies strongly
suggests that all of them contain an ILR \citep{sha93,fri96}.  Based
on existing data we cannot accurately and uniformly determine the
strength of the ILRs in these galaxies, so in this discussion
we assume that our sample contains galaxies with similar ILR
strengths. Table \ref{nucleustable} shows that with the exception of
NGC 4736 and NGC 5728, these galaxies share similar primary and
secondary bar sizes and ellipticities, so it is not unreasonable to
believe that the ILR strengths are similar in these galaxies. More
detailed observations at high resolution of the dynamics of these
galaxies are needed to confirm this, and thus validate the following
discussion.

The nuclear CO morphology of NGC 5728 resembles a clumpy structure
that is not centered on the galactic center. It appears that it may be
a lumpy circumnuclear ring that joins up with the southern-most dust
lane \citep{pet02}. High resolution CO maps of NGC 6951 show a ``twin
peaked'' structure that may be collections of molecular gas at the ILR
\citep{ken92,koh99}. NGC 470 and NGC 2273 show CO distributions
\citep{pet02,jog98} that coincide with the nuclear stellar bars seen
in the NIR images of \citet{woz95} and \citet{mul97}. The high
resolution CO maps of the inner 21\arcsec\ of NGC 4736 show an
elongated central peak which has been interpreted as a nuclear CO bar
by \citet{won00} and \citet{sak99}. We note that it is more centrally
concentrated than the bar-like features seen in CO in NGC 2273 and NGC
470.

There are currently no published high resolution CO maps of NGC 3081
and NGC 1097 (likely because of their low declination). As such, we
cannot include NGC 3081 in our comparison, but there are single dish CO
maps of NGC 1097 by \citet{ger88} that strongly suggest a circumnuclear
ring of CO, with a radius of $\sim$8\arcsec. This is confirmed by the
\ion{H}{1} maps by \citet{ond89}, so we will include this galaxy in our
comparisons.

The data in Table \ref{ratiostable} show evidence that galaxies
containing elongated molecular bar-like features in their central
regions have lower \twcott/\jto\ integrated intensity line
ratios. This effect is emphasized in Figure \ref{radial} where we show
the azimuthally averaged \twcooz\ radial profile compared with the
\twcott/\jto\ line ratio and the properties determined in
\S\ref{lte}. The profiles are ordered in decreasing value of the
\twcott/\jto\ line ratio. Clearly the degree of central concentration
increases as you reach the cooler gas near the bottom of the figure.
In the LVG approximation, the warmer and/or denser gas is at the top
of the plot, and temperature and/or density decreases downward. We
point out that there are resolution effects at work for the bottom
three galaxies (NGC 4736, NGC 470 and NGC 2273).  Since NGC 4736 is
the closest galaxy in our sample, we are seeing it at the highest
spatial resolution. Thus, in NGC 4736, our central peak appears much
narrower than in NGC 2273 and NGC 470, but in reality, the width of
each central peak is comparable to the resolution limit of the
interferometer with which they were observed. Despite this, there is
still an increasing tendency for the galaxies with lower \twcott/\jto\
line ratios to have more centrally concentrated gas distributions,
while the warmer (and possibly denser) gas appears to avoid the
nucleus. Possible causes for this effect are discussed in
\S\ref{disc1}.

\subsection{Comparisons with Recent Models \label{models}}

Previous models have indicated that in order to create and sustain
nuclear bars and rings, there must be some form of viscous gas
component that can dissipate energy \citep{com85,ath92,fri93,sha93}.
These models have reproduced many of the nuclear morphologies seen in
single- and double-barred galaxies by varying the model's parameters
such as bar pattern speed, main bar strength, ILR strength, \etc. Some
of these models have attempted to vary the gas viscosity parameter,
which is usually assumed to be some numerical viscosity or some
percentage of energy loss during cloud collision and kept constant for
all trials.  Our observational result is the first indication that
gas properties may be related to the variety of molecular gas
morphologies seen in the centers of double-barred galaxies.

Particularly relevant to this study are the studies of double-barred
systems by \citet{com94} and \cite{mac02}. \citet{com94} studied the
motions of molecular gas in a double-barred potential and found that
in galaxies with more viscous gas, the nuclear bar becomes
kinematically separate from the main bar \citep[as predicted
by][]{fri93}, while in the less viscous model, the nuclear bar is
actually rotating at the same pattern speed as the large-scale bar
\citep[as predicted by][]{sha93}. In both cases, the molecular gas in
the nucleus takes the form of a nuclear bar. We do not see these
nuclear gas bars in all of the CO maps, so it is difficult to compare
the observations with these models.

\citet{mac02} modeled the gas response to a double-barred potential
that could possibly be the result of purely stellar orbits. In this
model, the gas distribution need not resemble the stellar
distribution.  In their models, they find that the double-barred
potential is not capable of transporting molecular gas into the
nucleus with the efficiency predicted by \citet{shl89}. The gas still
becomes trapped at the ILR, unless they increase the sound speed. In
the simulations with higher sound speeds, the gas is capable of
flowing all the way into the nucleus. It is interesting to note that
our observations suggest that the galaxies with centrally concentrated
gas distributions contain cooler, less dense gas, which could
correspond to a higher sound speed. Thus, the gas in our double-barred
galaxies may be flowing inward more efficiently in the galaxies with
higher sound speeds as predicted by the models of \citet{mac02},
resulting in the more centrally concentrated CO profiles. However,
until we can accurately measure the sound speed, this conclusion must
remain rather speculative.

It seems more likely, though, that the variations in the gas
distributions are affected strongly by star formation activity, which is
not yet accounted for in these models. Future models of double-barred
galaxies will need to include the effects of star formation in order to
determine its effect on the flow of molecular gas into the centers of
double-barred galaxies.

\section{Speculations on the Origin of the Gas Profile - Gas Property Correlation \label{disc1}}

\subsection{The Effect of Star Formation on Gas Properties \label{sf}}

We were unable to find estimates of the star formation rates (SFR) of
all these galaxies at high resolution using the same technique. In an
attempt to keep the data sets used to compare the properties of the
galaxies in our sample as uniform as possible, we have estimated the
SFRs from the IRAS fluxes. While this may be the most accurate method
for these dusty star forming galaxy centers \citep{ken98}, the large
beam of the IRAS satellite means that in all cases we are likely
measuring the global SFR rather than the rates for the inner
21\arcsec\ (the beam size of the JCMT). For now, we will assume that
the majority of the star formation in these galaxies is occurring in
the central few kiloparsecs.

To estimate the SFR, we use the IRAS 60 $\mu$m and 100 $\mu$m fluxes
to determine the infrared luminosity as described in \citet{koh99}. We
then convert this luminosity to a star formation rate using the
technique described in \citet{hun86}. To test the accuracy of our
method, we compared the star formation rates for the entire galaxies
determined from the IRAS fluxes to the direct measurement for the star
formation rates using H$\alpha$ data for the inner 45\arcsec\ of a
sample of different galaxies in \citet{jog98}. We find that our star
formation rates are generally $\sim 2 - 3$ times higher than those
determined by \citet{jog98}, with somewhat larger discrepancies (up to
5 times higher) for the galaxies with low rates ($< 1$ \msun\
yr$^{-1}$).

In Figure \ref{sfr} we see that the \twcott/\jto\ line ratio tends to
be higher in the galaxies that show higher star formation rates. Given
that we are likely overestimating the star formation rates of NGC 4736
and NGC 2273 to a higher degree than we are for NGC 5728 and NGC 6951
(for reasons described in the previous paragraph) it is likely that
our result will hold when using a different technique to determine the
star formation rates.

It is interesting to note that the variation in the gas properties
with star formation activity discussed above, in conjunction with the
correlation noted in \S\ref{morph} (and shown in Figure \ref{radial})
indicates that the galaxies with higher star formation rates are
depleted of molecular gas in their centers. One possible cause could
be that the star formation and subsequent \ion{H}{2} regions and
supernovae deplete, photo-dissociate, and disrupt the clouds, leaving
the galaxies with higher star formation rates more gas deficient in
the nucleus, while heating and compressing the surrounding molecular gas.
Without uniform, high resolution observations of the star formation
rates and locations, we cannot draw any firm conclusions from this
analysis.

\subsection{The Effect of Gas Properties on Star Formation \label{dyn}}

In the previous section we assumed that the dynamics in these galaxies
were similar based on the similarity of the NIR images. There are
several unmeasured quantities that can change the central gas dynamics
of these galaxies. One example is the bar pattern speed. Changing the
pattern speed, will change the strength of the ILR.  We cannot rule
out the possibility that the variation in gas distributions are caused
by different strength ILRs. If this is the case, then the variations
in star formation activity may be caused by the locations of the peaks
in the gas distributions.  For example, if the majority of gas is
located at the ILR radius (a local stable point) it may be able to
accumulate until it exceeds the critical density needed for star
formation to occur \citep[see, for example][and references
therein]{elm94}. Conversely, if the ILR is weak, the majority of the
gas could accumulate in the central regions of the galaxy, where the
local epicycle frequency \citep[hence, critical density;][and
references therein]{ken89} may be high enough to inhibit strong star
formation in these galaxies. This may result in higher star formation
rates in galaxies like NGC 6951 and NGC 1097 where the gas is
concentrated near the ILR \citep{koh99,ger88} than those galaxies with
strong central concentration such as NGC 4736 \citep{won00}.

A detailed study of a larger sample of galaxies combined with a more
detailed analysis of the star formation histories and dynamics of
these galaxies will help clarify this picture among double-barred
galaxies.

\section{Summary \label{conc}}

We have taken \twcott, \thcott, and \twcott\ spectra of the ILR
regions of a sample of galaxies that show a double-barred nucleus in
the near infrared. We find that that \twco/\thcoto\ integrated
intensity line ratio is uniform at the 2$\sigma$ level while the
\twcott/\jto\ line ratio varies dramatically from galaxy to galaxy.

We find that the CO \jtt/\jto\ line ratio is higher in galaxies with
non-centralized CO distributions and lower in galaxies that contain
centralized concentrations. The uniform \twco/\thcoto\ line ratios
suggest that the molecular gas is marginally optically thick, with
$\tau \sim 5$ in all galaxies. The similarity in optical depths
suggests that the variation in the \twcott/\jto\ line ratio is
indicating real variations in the gas properties. This result suggests
that the molecular gas is cooler and/or less dense in galaxies that
contain central concentrations and warmer and/or denser in galaxies
with less centrally concentrated CO distributions.

There is some evidence that the star formation rates are higher in the
galaxies that have warmer/denser, non-centrally concentrated gas
distributions. This result suggests that the star formation activity
may be depleting the central gas while heating and compressing the
surrounding molecular gas resulting the the observed variety of gas
distributions and physical conditions. 

It is also possible that bar dynamics alone is responsible for
modifying the gas distributions, densities, and star formation rates.
In this scenario the enhanced star formation is the result of stronger
ILRs allowing molecular gas to accumulate to high (super-critical)
densities near the ILR radius. A more detailed study of the star
formation activity and internal dynamics of these galaxies at higher
resolution is needed to disentangle these.

It seems evident based on these observations that future models of
double-barred galaxies will need to include star formation in order to
disentangle the effects of dynamics and star formation activity.

\acknowledgments 

This research has been supported by a research grant to C.~D.~W.~from
NSERC (Canada). G.~R.~P.~is supported by NSF grant AST 99-81289 and by
the State of Maryland via support of the Laboratory for
Millimeter-Wave Astronomy. This research has made use of the NASA/IPAC
Extragalactic Database (NED) which is operated by the Jet Propulsion
Laboratory, California Institute of Technology, under contract with
the National Aeronautics and Space Administration. We wish to thank
S.~Jogee for providing the CO data for NGC 6951 and NGC 470 from her
Ph.D.  thesis and T.~Wong for providing the data for NGC 4736 so that
we could calculate the radial profiles. We also wish the thank the
referee Eric Emsellem for detailed comments that greatly improved
the clarity and quality of this paper.

\clearpage

\figcaption[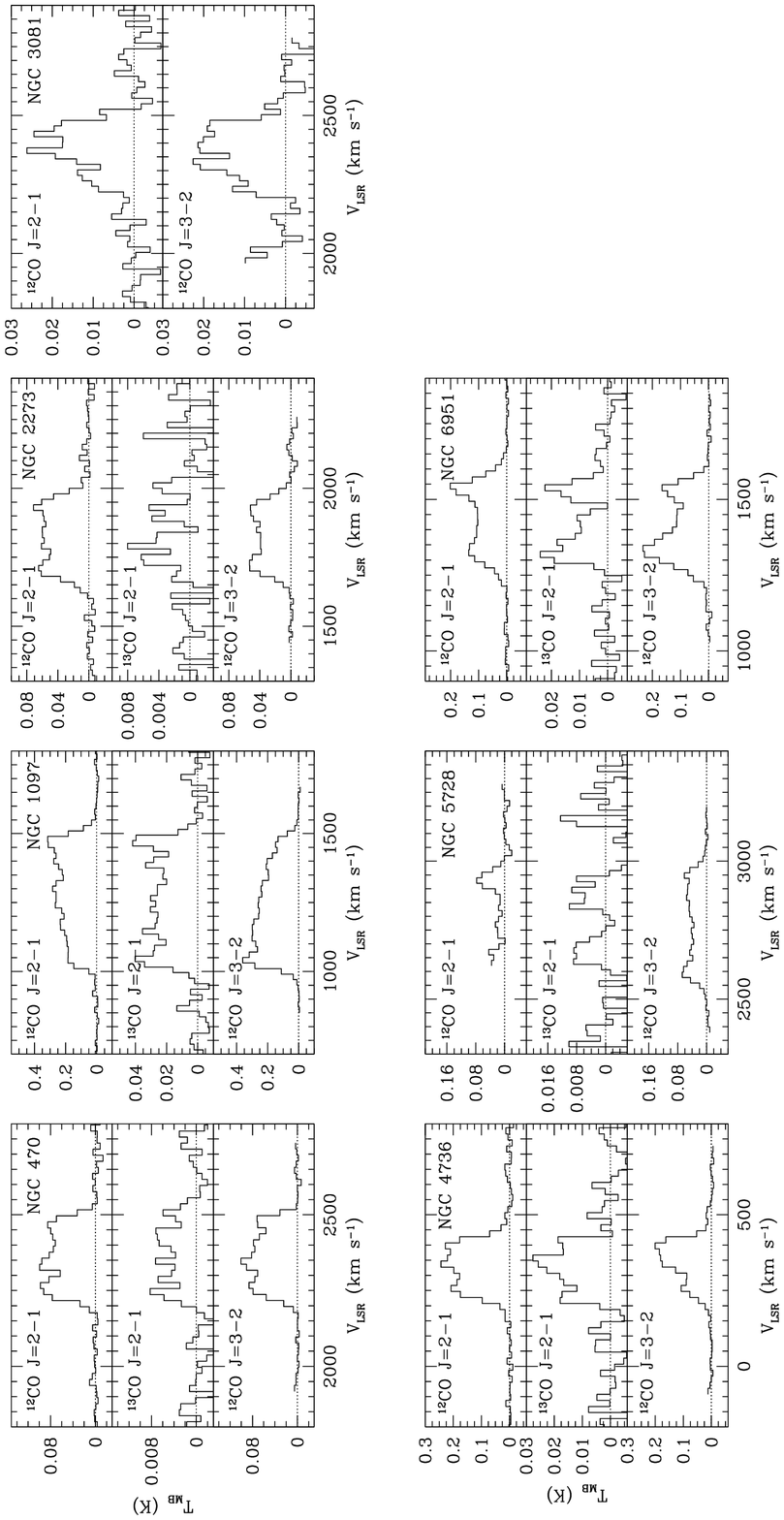]{Individual spectra for the inner 21$''$ of
each galaxy. The top panel shows \twcoto, the middle panel shows
\thcoto, and the bottom panel shows \twcott\ (convolved to 21$''$
resolution).  The vertical scales for the \twcoto\ and \twcott\
spectra are the same, while the scale of the \thcoto\ spectra is
10$\times$ smaller for each galaxy.
\label{spectra}
}

\figcaption[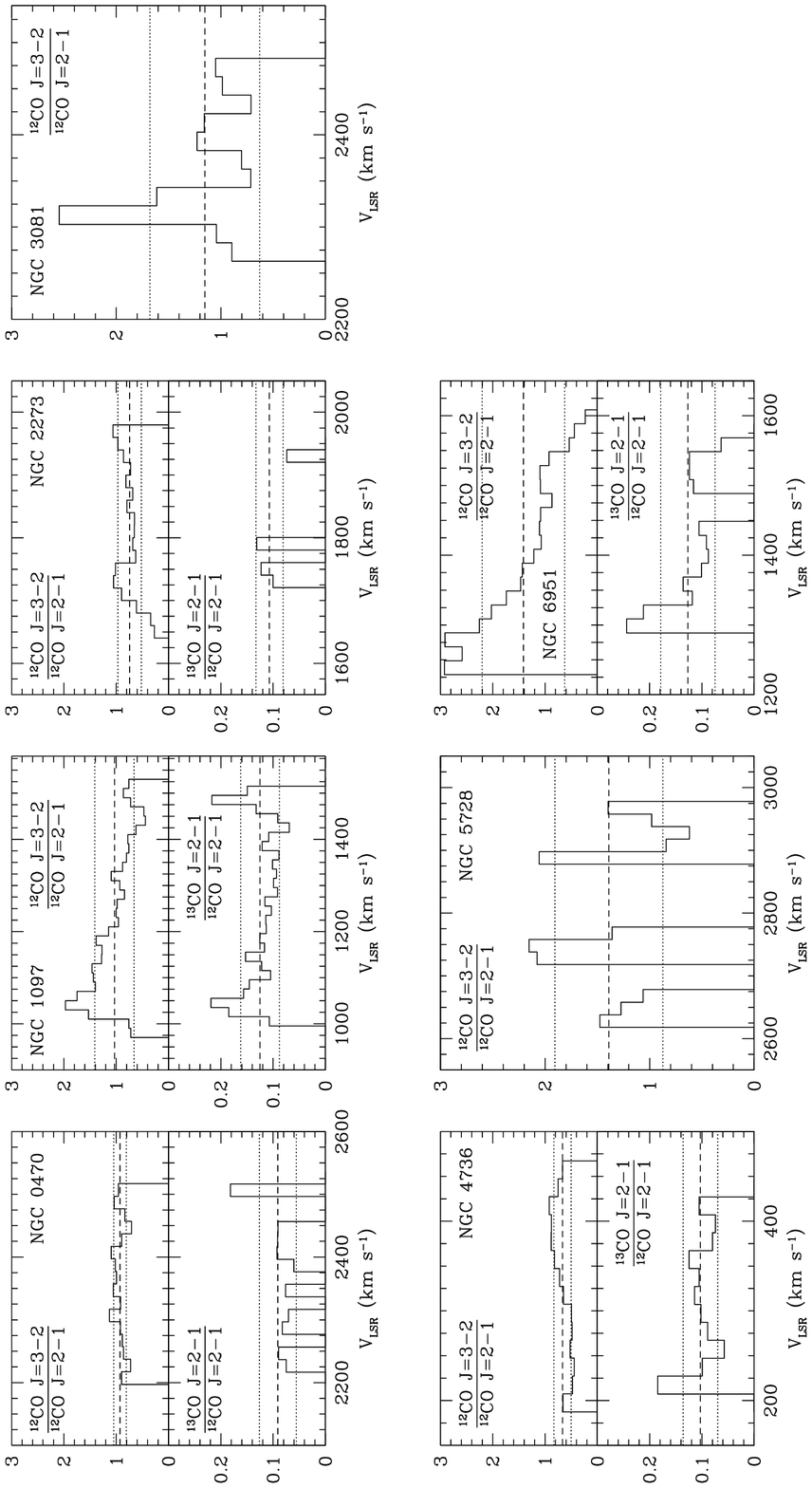]{Channel by channel line ratios for each galaxy.
The top panel shows the \thco/\twcoto\ ratio and the bottom panel
shows the \twcott/\jto\ ratio. Only those channels with a combined SNR
$>$ 2 are shown. The dashed line is the average over all channels, the
dotted line is the root-mean-square of the scatter. The values for the
ratios listed in Table \ref{ratiostable} are the average value over
all channels shows here, and the uncertainty is the standard deviation
in the mean. For all galaxies, the vertical scale ranges from 0 to 3
for the \twcott/\jto\ line ratio and 0 to 0.3 for the \thco/\twcoto\
line ratio.
\label{ratios}
}

\figcaption[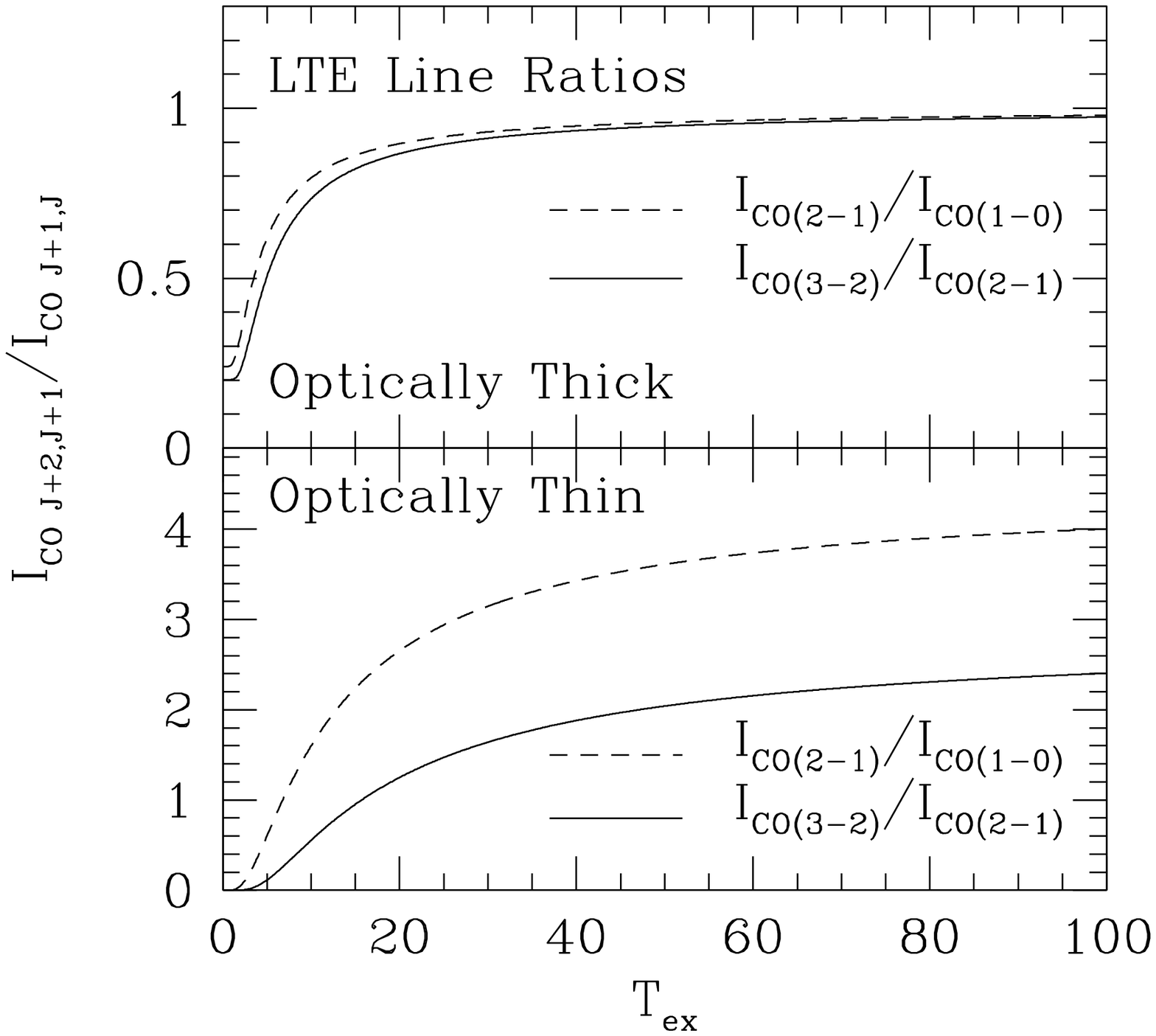]{Plots of \twco\ integrated intensity line
ratio as a function of excitation temperature in the LTE approximation.
\label{lteplots} 
}

\figcaption[f4.eps]{Plots of CO column density (per unit velocity)
versus H$_2$ density for a kinetic temperature of 30 K and a
\twco/\thco\ abundance ratio of 50. The contours represent the $\pm$
30\% contours for the CO line ratios given in Table \ref{ratiostable}
and are indicated in the lower left panel. The underlying grayscale is
the \twcott/\jto\ line ratio. For NGC 4736 we have included the
\twco/\thcooz\ ratio of \citet{sag91}. Solutions appear as regions
where all line ratios overlap. In the case where only one contour is
visible on the plot, an arrow indicates the allowed regions.
\label{lvgplots} 
}

\figcaption[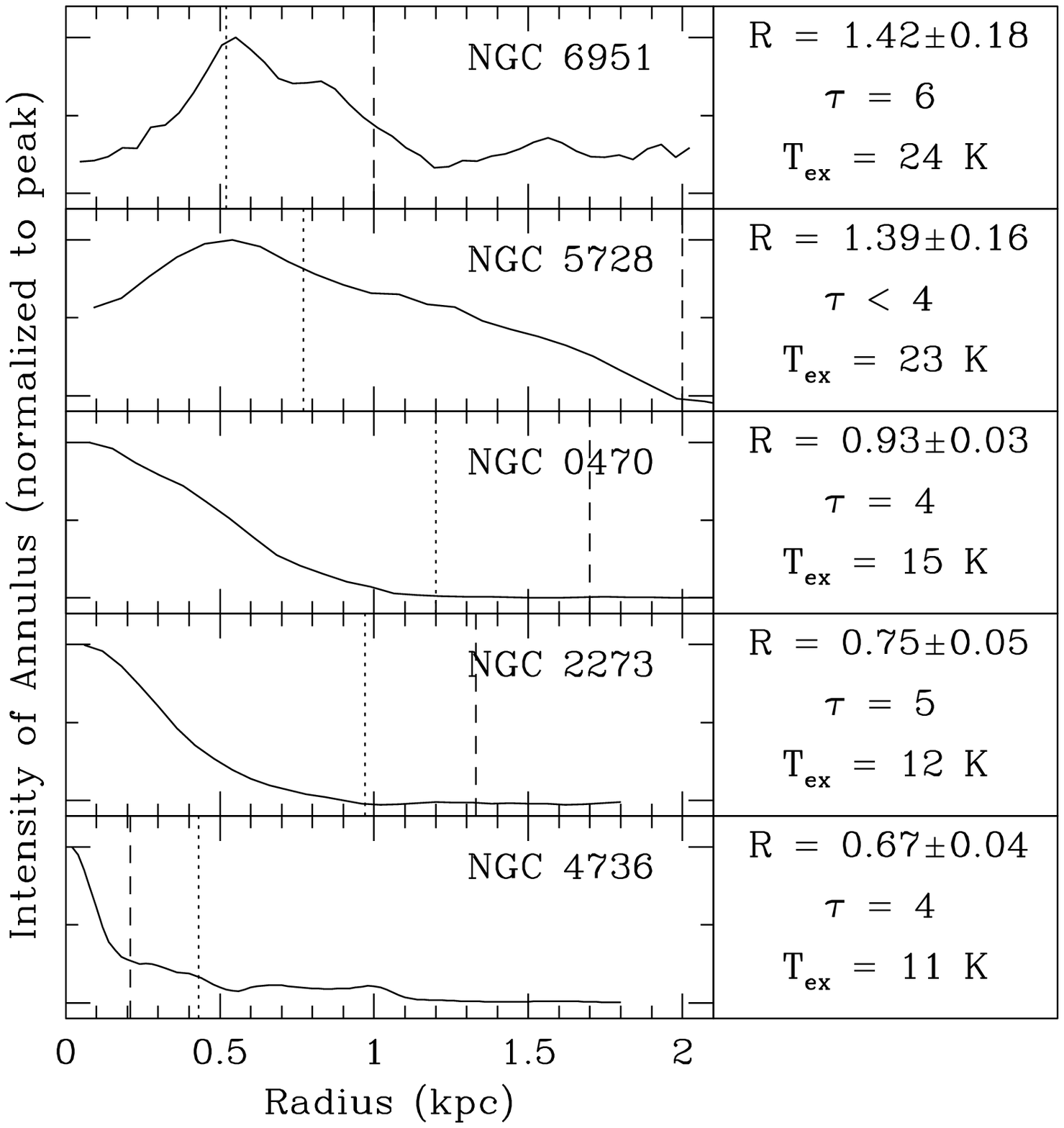]{Radial profiles of the CO intensity
compared to the \twcott/\jto\ line ratio. The solid line is the
azimuthally averaged CO distribution from high resolution
observations.  The dotted line shows the location of the ILR as traced
by the ends of the nuclear stellar bar and star forming rings
\citep[when present;][]{fri96,woz95,mul97}. The dashed line represents
the area covered by the JCMT primary beam. Also shown are the
excitation temperatures and optical depths assuming LTE. They are
arranged such that the galaxy with the highest line ratio is at the
top, and the line ratio decreases as you move down.  Note that the
radial profiles become increasingly centrally concentrated as the line
ratio decreases. Note that the bottom three galaxies are resolution
limited.
\label{radial}
}

\figcaption[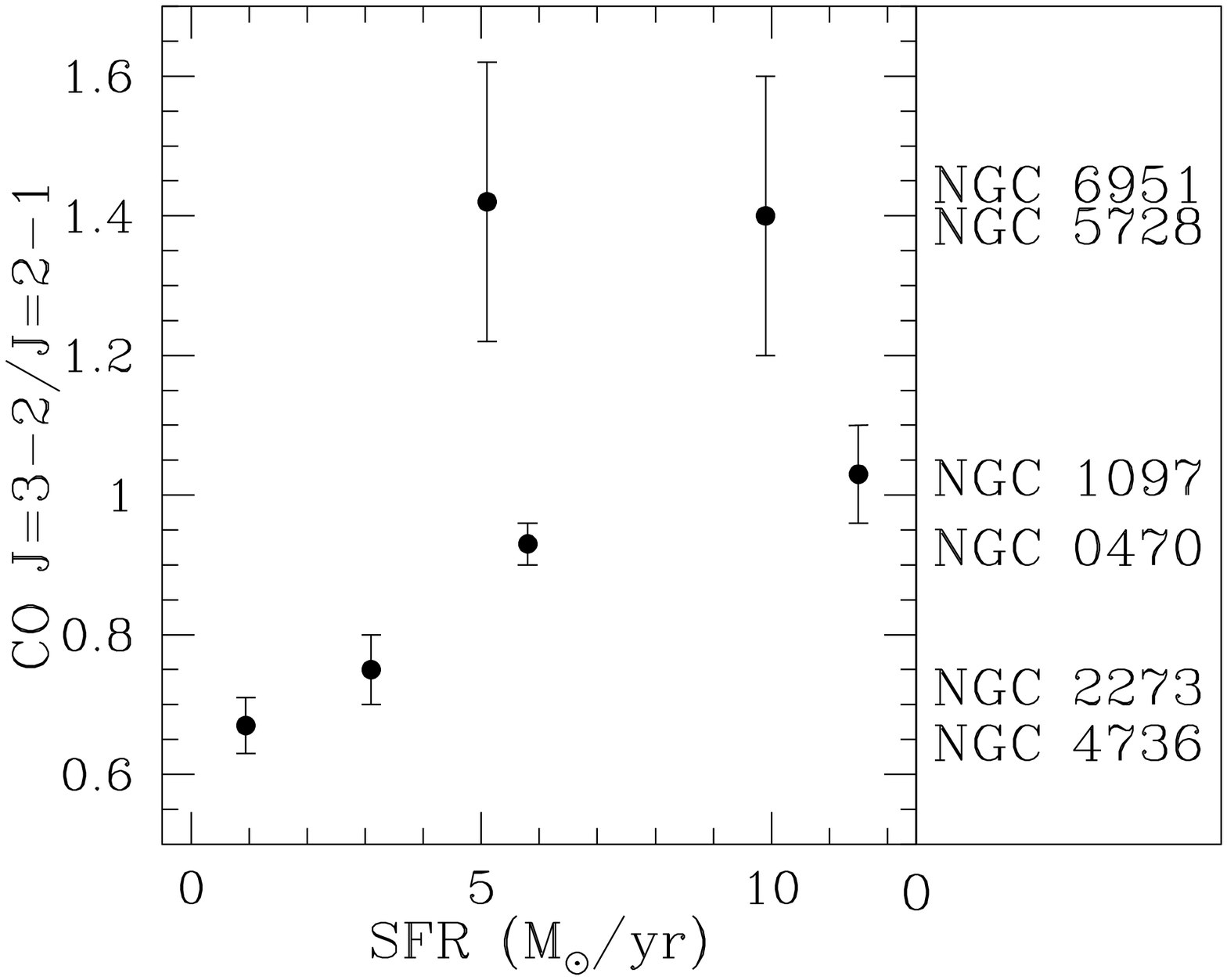]{Comparison of (global) star formation rates
to (nuclear) \twcott/\jto\ line ratio. Since the optical depths are similar in
each galaxy, the CO line ratio suggests that the gas is warmer and/or
denser in the galaxies with higher star formation rates determined from
IRAS fluxes. Higher resolution tracers of star formation activity and
rates are needed to confirm this finding.
\label{sfr} 
}

\begin{deluxetable}{lcccccl}
\tablecolumns{7}
\tablewidth{0pc}
\tablecaption{Galaxy Properties \label{galaxytable}}
\tablehead{
\colhead{Galaxy} & 
\colhead{RA} &
\colhead{Dec} &
\colhead{$V_{\rm LSR}$} &
\colhead{d} &
\colhead{RC3} &
\colhead{Nuclear} \\
\colhead{~} & 
\colhead{(2000)} &
\colhead{(2000)} &
\colhead{(\kms) } &
\colhead{(Mpc)} &
\colhead{Class.} &
\colhead{Activity}          } 
\startdata
NGC 0470 & 01\h19\m44\fs8 & $+$03\arcdeg24\arcmin35\arcsec & 2374 & 32 & SA(rs)b            & starburst  \\
NGC 1097 & 02\h46\m19\fs0 & $-$30\arcdeg16\arcmin29\arcsec & 1275 & 17 & (R$'_1$)SB(r$'$l)b & Seyfert 1 \\
NGC 2273 & 06\h50\m08\fs7 & $+$60\arcdeg50\arcmin45\arcsec & 1871 & 25 & SB(r)a             & Seyfert 2 \\
NGC 3081 & 09\h59\m29\fs5 & $-$22\arcdeg49\arcmin35\arcsec & 2385 & 32 & (R$_1$)SAB(r)0/a   & Seyfert 2 \\
NGC 4736 & 12\h50\m53\fs0 & $+$41\arcdeg07\arcmin14\arcsec &  308 &  4 & (R)SA(r)ab         & LINER     \\
NGC 5728 & 14\h42\m23\fs9 & $-$17\arcdeg15\arcmin10\arcsec & 2788 & 37 & (R$_1$)SAB(r)a     & Seyfert 2 \\
NGC 6951 & 20\h37\m14\fs5 & $+$66\arcdeg06\arcmin20\arcsec & 1424 & 19 & SAB(rs)bc          & Seyfert 2 \\
\enddata
\tablecomments{All data taken from the NASA IPAC Extragalactic
Database. Distances are derived assuming H$_0$ = 75 \kms\ kpc$^{-1}$.}
\end{deluxetable}

\begin{deluxetable}{lcccccl}
\tablecolumns{6}
\tablewidth{0pc}
\tablecaption{Primary and Secondary Bar Properties \label{nucleustable}}
\tablehead{
\colhead{Galaxy} & 
\colhead{$L_1$} &
\colhead{$\epsilon_1$} &
\colhead{$L_2$} &
\colhead{$\epsilon_2$} &
\colhead{Refs} \\
\colhead{~} & 
\colhead{(kpc)} &
\colhead{~} &
\colhead{(kpc)} &
\colhead{~} &
\colhead{~}          } 
\startdata
NGC 0470 &   4.6  &     0.55    & 1.2 & 0.42 - 0.46 & 1,2  \\
NGC 1097 &   6.6  &     0.67    & 0.9 & 0.42 - 0.46 & 1,2  \\
NGC 2273 &   2.9  &     0.37    & 1.0 &     0.28    & 3    \\
NGC 3081 &   6.0  & 0.65 - 0.66 & 1.6 & 0.37 - 0.42 & 1,2  \\
NGC 4736 &\nodata &   \nodata   & 0.6 &  $\sim$0.3  & 4     \\
NGC 5728 &  13.2  &     0.71    & 0.8 &     0.49    & 1    \\
NGC 6951 &   4.5  & 0.5 - 0.6   & 0.5 &     0.31    & 1,2  \\
\enddata
\tablecomments{$L_1$ and $L_2$ are the lengths of the primary and
secondary bars (respectively) while $\epsilon_1$ and $\epsilon_2$ are
the ellipticities of the primary and secondary bars
(respectively). Physical sizes are derived assuming H$_0$ = 75 \kms\
kpc$^{-1}$.}
\tablerefs{(1) \citet{woz95}; (2) \citet{fri96}; (3) \citet{mul97}; (4) \citet{mol95}.}
\end{deluxetable}

\begin{deluxetable}{lcccccl}
\tablecolumns{7}
\tablewidth{0pc}
\tablecaption{Spectral Line Parameters \label{spectratable}}
\tablehead{
\colhead{Galaxy} & 
\colhead{$^{12}$CO J=2-1} &
\colhead{$^{13}$CO J=2-1} &
\colhead{$^{12}$CO J=3-2} &
\colhead{$^{12}$CO J=3-2} &
\colhead{$^{12}$CO J=1-0} &
\colhead{Refs} \\
\colhead{~} & 
\colhead{(21\arcsec)} &
\colhead{(21\arcsec)} &
\colhead{(21\arcsec)} &
\colhead{(16\arcsec)} &
\colhead{~} &
\colhead{~}           } 
\startdata
NGC 0470 & 24.2 $\pm$ 0.5 &  1.8 $\pm$ 0.2 & 22.4 $\pm$ 0.3 & 24.9 $\pm$ 0.6 & 19.3 (16\arcsec) & 1 \\
NGC 1097 &  121 $\pm$ 1   & 14.6 $\pm$ 0.6 &121.7 $\pm$ 0.5 &137.5 $\pm$ 0.8 & 80 (16\arcsec)   & 2 \\
NGC 2273 & 19.7 $\pm$ 0.5 &  1.2 $\pm$ 0.2 & 14.3 $\pm$ 0.3 & 17.9 $\pm$ 0.5 & 22.5 (22\arcsec) & 3 \\
NGC 3081 &  4.6 $\pm$ 0.3 &        ...     &  4.9 $\pm$ 0.3 &  7.4 $\pm$ 0.5 & ...   & ... \\
NGC 4736 & 46.1 $\pm$ 0.8 &  4.6 $\pm$ 0.4 & 31.9 $\pm$ 0.4 & 32.9 $\pm$ 0.5 & 32.5 (16\arcsec) & 4 \\
NGC 5728 &  9.5 $\pm$ 0.6\tablenotemark{1} & (0.8 $\pm$ 0.8)& 21.5 $\pm$ 0.4 & 26.5 $\pm$ 0.8 & 9.74 (16\arcsec) & 2 \\
NGC 6951 & 41.7 $\pm$ 0.6 &  4.2 $\pm$ 0.4 & 48.8 $\pm$ 0.5 & 59.4 $\pm$ 0.8 &  71 (16\arcsec)  & 5 \\
\enddata
\tablenotetext{1}{The \twcoto\ flux for NGC 5728 may be underestimated due to the poor centering of the spectrometer.}
\tablecomments{All CO line strengths are in K \kms\ in \tmb. The beam sizes for the first
four columns are stated below the transition. The \twcooz\ beam sizes are given beside the flux value.
Observations of \thcoto\ in NGC 3081 were not attempted. \\
\vskip 0.1 cm
\hskip 20pt
\twcooz\ References. --- (1) \citet{sof93}; (2) \citet{vil98}; (3) \citet{kru90}; (4) \citet{shi98}; (5) \citet{koh99} }
\end{deluxetable}

\begin{deluxetable}{lcccccl}
\tablecolumns{7}
\tablewidth{0pc}
\tablecaption{CO Line Ratios and Nuclear Characteristics \label{ratiostable}}
\tablehead{
\colhead{Galaxy} & \colhead{{$^{13}{\rm CO} J=2-1\over{^{12}{\rm CO} J=2-1}$}\tablenotemark{(a)}} & \colhead{{$^{12}{\rm CO} J=3-2\over{^{12}{\rm CO} J=2-1}$}\tablenotemark{(a)}} &\colhead{{$^{12}{\rm CO} J=2-1\over{^{12}{\rm CO} J=1-0}$}\tablenotemark{(b)}} & \colhead{Nucl. CO}& \colhead{Activity\tablenotemark{(c)}}& \colhead{Refs} \\
          } 
\startdata
NGC 0470 & 0.09$\pm$0.01 & 0.93$\pm$0.03 & 1.38$\pm$0.42 &  elongated    &  SB & 1,2     \\
NGC 1097 & 0.13$\pm$0.01 & 1.03$\pm$0.07 & 1.67$\pm$0.50 &  ring?  &  S1 & 3,4     \\
NGC 2273 & 0.11$\pm$0.01 & 0.75$\pm$0.05 & 0.88$\pm$0.26 &  elongated    &  S2 & 3,5,6   \\
NGC 3081 &      ...      &  1.2$\pm$0.2  &    ...        &  ?      &  S2 & 7       \\
NGC 4736 & 0.10$\pm$0.01 & 0.67$\pm$0.04 & 0.70$\pm$0.14 & bar/peak &  L  & 8,9  \\
NGC 5728 &    $<$ 0.09   &  1.4$\pm$0.2  & 1.96$\pm$0.59 & irregular &  S2 & 3,6     \\
NGC 6951 & 0.13$\pm$0.02 &  1.4$\pm$0.2  & 0.59$\pm$0.18 &  twin peaks &  S2 & 3,10,11 \\
\enddata
\tablenotetext{(a)}{All data for these columns are taken with the
JCMT. The line ratio calculation is described in the text.}
\tablenotetext{(b)}{The \twcooz\ data for NGC 2273 is taken with the
IRAM 30 m telescope with a 22\arcsec\ beam. The line ratio for NGC 4736
is taken from \citet{ger91} with a 22\arcsec\ beam. The remaining
ratios are a combination of \twcott\ data taken with the JCMT
(16\arcsec\ beam) and \twcooz\ data taken with the Nobeyama 45 m
telescope (16\arcsec\ beam). The \twcott/\joz\ ratio is scaled to the
\twcoto/\joz\ ratio by dividing by the \twcott/\jto\ ratio in column
3. All ratios are given in \tmb.}
\tablenotetext{(c)}{SB = Starburst; S1 = Seyfert 1; S2 = Seyfert 2; L =
LINER} \tablerefs{(1) \citet{sof93}; (2) \citet{jog98}; (3)
\citet{vil98}; (4) \citet{ger88}; (5) \citet{kru90}; (6)
\citet{pet02}; (7) \citet{phi83}; (8) \citet{ger91}; (9)
\citet{sak99}; (10) \citet{aal95}; (11) \citet{ken92} }

\end{deluxetable}

\begin{deluxetable}{lcccccl}
\tablecolumns{7}
\tablewidth{0pc}
\tablecaption{LTE Parameters and Dust Temperatures \label{ltesols}}
\tablehead{
\colhead{Galaxy} & \colhead{T$_{ex}^{J=3-2}$} & \colhead{T$_{ex}^{J=3-2}$} &\colhead{$\tau_{21}$} &\colhead{f$_{60}$/f$_{100}$} &\colhead{T$_{dust}$} & \colhead{Refs.} \\
\colhead{~} & \colhead{(thin)} & \colhead{(thick)} &\colhead{~} &\colhead{~} &\colhead{($\beta = 2,1$)} & \colhead{~}          } 
\startdata
%        & 32thin  & 32thick  & tau    & f60  & Tdust    & refs
NGC 0470 & 15 K    & $>$ 38 K & 3.9    & 0.51 & 30, 37 K & 1  \\
NGC 1097 & 16 K    & $>$ 66 K & 5.6    & 0.52 & 30, 36 K & 1  \\
NGC 2273 & 12 K    & 11 K     & 4.7    & 0.60 & 31, 38 K & 1  \\
NGC 3081 & 18 K    & $>$ 150 K& ...    &  ... & ...      &    \\
NGC 4736 & 11 K    & 8 K      & 4.2    & 0.57 & 31, 37 K & 2  \\
NGC 5728 & 23 K    & ...      & $<$ 3.9& 0.60 & 31, 38 K & 2  \\
NGC 6951 & 24 K    & ...      & 5.6    & 0.35 & 26, 31 K & 1  \\
\enddata

\tablecomments{For the optically thick column, we assumed the
$-1\sigma$ limit for line ratios that were consistent with 1. The
missing entries indicated that the line ratios were not consistent with
optically thick emission within the uncertainties. For $\tau$ we assume
a \twco/\thco\ abundance ratio of 43 \citep{haw87}. For T$_{dust}$ the
two values represent a dust emissivity ($\beta$) of 2 or 1
respectively \citep{dev90}.
}
\tablerefs{(1) \citet{iras}; (2) \citet{soi89} }

\end{deluxetable}

\begin{deluxetable}{lccccc}
\tablecolumns{6}
\tablewidth{0pc}
\tablecaption{LVG Solutions for a Kinetic Temperature of 30 K \label{lvgsols}}
\tablehead{
\colhead{Galaxy} & \colhead{log N(CO)/dv} & \colhead{log n(H$_2$)} &\colhead{$\tau_{10}$} &\colhead{$\tau_{21}$} &\colhead{$\tau_{32}$}  \\
          } 
\startdata
NGC 0470  &  16.8 $\pm$ 0.3  & $>$ 3.7       & $<$ 1.6 & $<$ 5.4  &  $<$ 7.4  \\
NGC 1097  &  17.0 $\pm$ 0.1  & $>$ 4.5       & $<$ 2.3 & $<$ 7.2  &  $<$ 10   \\
NGC 2273  &  18.0 $\pm$ 1.1  & unbounded     & $<$ 64  & $<$ 190  &  $<$ 190  \\
NGC 4736  &  17.0 $\pm$ 0.1  & 3.1 $\pm$ 0.3 & 4.9     &    14    &  19       \\
NGC 5728  &  $<$ 16.5        & $>$ 4.6       & $<$ 0.5 & $<$ 1.9  &  $<$ 2.7  \\
NGC 6951\tablenotemark{1}  &  $<$ 16.5        & $>$ 4.6       & $<$ 0.5 & $<$ 1.9  &  $<$ 2.7  \\
\enddata
\tablenotetext{1}{Constraints are based only on the \twcott/\jto\ line ratio.}
\end{deluxetable}

\clearpage

\plotone{f1.eps}
\vfil
\plotone{f2.eps}
\vfil
\plotone{f3.eps}
\vfil
\plotone{f4.eps}
\vfil
\plotone{f5.eps}
\vfil
\plotone{f6.eps}

\end{document}